
\headline={\ifnum\pageno=1\firstheadline\else
\ifodd\pageno\rightheadline \else\leftheadline\fi\fi}
\def\firstheadline{\hfil}
\def\rightheadline{\hfil}
\def\leftheadline{\hfil}
        \footline={\ifnum\pageno=1\firstfootline\else\otherfootline\fi}
\def\firstfootline{\rm\hss\folio\hss}
\def\otherfootline{\hfil}

\font\tenrm=cmr10

\font\elevenbf=cmbx10 scaled\magstep 1
\font\elevenrm=cmr10 scaled\magstep 1
\font\elevenit=cmti10 scaled\magstep 1

\font\ninerm=cmr9

\nopagenumbers
\hsize=6.0truein
\vsize=8.5truein
\parindent=1.5pc
\baselineskip=10pt
\centerline{\elevenbf SUSY QCD CORRECTIONS TO THE $t\rightarrow H^+b$\
DECAY\footnote*{{\ninerm\baselineskip=11pt
Talk presented at MRST--94, McGill U.,
Montr\'eal, Canada, May 11--13 and
SUSY--94, U. of Michigan, Ann Arbor, USA, May 14--17, UQAM--PHE--94/04.}}}
\vglue 7pt
\vglue 1.0cm
\centerline{\elevenrm HEINZ K\"ONIG}
\baselineskip=12pt
\centerline{\elevenit D\'epartement de Physique}
\baselineskip=12pt
\centerline{\elevenit Universit\'e du Qu\'ebec \`a Montr\'eal}
\baselineskip=12pt
\centerline{\elevenit C.P. 8888, Succ. Centre Ville, Montr\'eal}
\baselineskip=12pt
\centerline{\elevenit Qu\'ebec, Canada H3C 3P8}
\vglue 0.8cm
\centerline{\tenrm ABSTRACT}
\vglue 0.3cm
  {\rightskip=3pc
 \leftskip=3pc
 \tenrm\baselineskip=12pt
 \noindent
I present the contribution of gluinos and scalar quarks to
the decay rate of the top quark into a charged Higgs boson and
a bottom quark within the minimal supersymmetric standard
model, including the mixing of the scalar partners of
the left- and right-handed
 top quark. I show that for certain values of the
supersymmetric parameters the standard QCD loop corrections
to this decay mode are diminished or enhanced by several
tens of per cent. I show that not only
the small value of 3 GeV for the
gluino mass (small mass window) but also much larger
values of several hundreds of GeV's
have a non-neglible effect
on this decay rate, against general belief.
 Last but not least, if the ratio of the
vacuum expectation values of the Higgs bosons are taken
in the limit of $v_1\ll v_2$ I obtain a drastic
enhancement due to a $\tan\beta$\ dependence in the couplings.
\vglue 0.8cm }
\line{\elevenbf I. INTRODUCTION\hfil}
\vglue 0.5cm
\baselineskip=14pt
\elevenrm
Recently there has been a lot of interest in the
electroweak loop corrections$^{1,2}$\
as well as in the QCD loop corrections$^{3,4,5,6}$\
to the top quark decay into a charged Higgs boson
and a bottom quark.\hfill\break\indent
In the standard model we have no charged Higgs particle
and therefore this decay can be used as a test for models
beyond the standard model; such as a two Higgs
doublet model$^7$\
and the minimal supersymmetric
extensions of the standard model (MSSM)$^{8,9}$, which is the
 favorite model beyond the standard model.
\hfill\break\indent
In this talk I take the last one as the underlying
model to consider the QCD corrections to the $t\rightarrow H^+b$\
decay mode. Czarnecki and Davidson$^6$\ showed that the effect
of the mass of the bottom quark to this decay rate is negligible.
They also showed that the ratio of the first
order to the zeroth order is constant at about $-9\%$ for a
wide range of the Higgs mass ($0\le m_{H^+}\le 90$\ GeV); the
top quark mass was taken to be $150$\ GeV.\hfill\break\indent
In this talk I present the SUSY QCD loop corrections to the
 $t\rightarrow H^+b$\ decay if gluinos and scalar quarks
are taken within the loop. Throughout
the calculations I neglect the mass of the bottom quark,
but I do not neglect the mixing of the scalar partners of
the left- and right-handed
top quark, which is proportional to the top quark
mass.\hfill\break\indent
In the next section I only present the results and refer the
interested reader for detailed calculation to ref.10. I end
with final remarks and the conlusions.
\vglue 0.65cm
{\elevenbf II. SUSY QCD CORRECTIONS TO THE TOP QUARK DECAY INTO
A CHARGED HIGGS BOSON AND BOTTOM QUARK}
\vglue 0.65cm
In the MSSM
the interaction Lagrangian relevant to the
decay $t\rightarrow H^+b$\
leads to the following decay rate
for $m_b\ll m_{\rm top}$:
$$\Gamma^0(t\rightarrow H^+b)={{G_F}\over{\sqrt{2}}}\vert
V_{tb}\vert^2\cot^2\beta{1\over{8\pi}}m_{\rm top}^3\bigl (1-
{{m_{H^+}^2}\over{m^2_{\rm top}}}\bigr)^2\eqno(1)$$
$V_{tb}\approx 1$\ is the Kobayashi-Maskawa matrix value,
$\cot\beta=v_1/v_2$\ is the ratio of the
vacuum expectation values (vev) of the two Higgs doublets.
\hfill\break\indent
To calculate the 1 loop diagram
with gluinos and scalar quarks within the loop we need the
couplings of the scalar quarks to the charged Higgs boson
and the scalar-quark-gluino-quark coupling. The first coupling
is given in Fig.115\footnote*{
\ninerm\baselineskip=11pt $\mu$\ has to be replaced by $-\mu$}
in ref.7 and the latter one in Eq.(C89) in ref.9.\hfill\break\indent
When neglecting the bottom quark mass only the scalar partner of
the left handed bottom quark $\tilde b_L$\ occurs within the loop whereas
for the top quark we have to take both left- and right-handed
superpartner $\tilde t_L$\ and $\tilde t_R$\ into account.
Furthermore since the mixing of $\tilde t_L$\ and $\tilde t_R$\
is proportional to the top quark mass we have to include the
full scalar top quark matrix, which is given by:
$$M^2_{\tilde t}=\left(\matrix{m^2_{\tilde t_L}
+m_{\rm top}^2+0.35D_Z^2&
-m_{\rm top}(A_{\rm top}+\mu\cot\beta)\cr-m_{\rm top}
(A_{\rm top}+\mu\cot\beta)&
m^2_{\tilde t_R}+m^2_{\rm top}+0.16D_Z^2\cr}\right)\eqno(2)$$
where $D_Z^2=m_Z^2\cos 2\beta$.
$m^2_{\tilde t_{L,R}}$\ are soft breaking masses,
 $A_{\rm top}$\ is the trilinear
scalar interaction parameter and $\mu$\ is the supersymmetric
mass mixing term of the Higgs bosons.
The mass eigenstates
$\tilde t_1$\ and $\tilde t_2$\ then are related to the current
eigenstates $\tilde t_L$\ and $\tilde t_R$ by
${\displaystyle{\tilde t_1=cos\Theta\tilde
t_L+\sin\Theta\tilde t_R\qquad
\tilde t_2=-\sin\Theta\tilde t_L+\cos\Theta\tilde t_R}}$.
 In the following we take
$m_{\tilde t_L}=m_{\tilde t_R}=m_S=A_{\rm top}$\
(global SUSY), $m^2_{\tilde b_1}=m_S^2
-0.42D^2_Z$\ and $m^2_{\tilde b_2}=m_S^2
-0.08D^2_Z$.
With negelcting bottom quark mass
the scalar partners of the left and right handed bottom quarks
do not mix and therefore $m_{\tilde b_L}=
m_{\tilde b_1}$.
 The gluino mass $m_{\tilde g}$\ is a free parameter,
which in general is supposed to be larger than 100 GeV, although
there is still the possibility of a small gluino mass window in
the order of 1 GeV$^{11,12}$.\hfill\break\indent
In Eq.(10) in ref.6 the authors present the results of the
standard QCD 1 loop corrections within the two Higgs doublet
model and the MSSM, which I will include in my calculation.
The results of the calculation of the loop diagram with
scalar quarks and gluinos
are finite, there are no dimensional divergencies.
As a result
I get for the first order in $\alpha_s$:
$$\eqalignno{\Gamma^1(t\rightarrow H^+b)=&\Gamma^0(t\rightarrow
H^+b)\Bigl\lbrack 1+{{4\alpha_s}\over{3\pi}}\tilde G'_+-
{{2\alpha_s}\over{3\pi}}(S+A)\Bigr\rbrack&(3)\cr
\tilde G'_+=&2{\rm Li}_2(1-\chi^2)-{{\chi^2}\over{1-\chi^2}}
\log{\chi^2}+\log{\chi^2}\log{(1-\chi^2)}+{1\over{\chi^2}}
(1-{5\over 2}\chi^2)\cdot\cr
&\log{(1-\chi^2)}
-{{2\pi^2}\over{3}}+{9\over 4}\cr
\chi^2=&{{m_{H^+}^2}\over{m_{\rm top}^2}}\cr
S=&S_t+{{m_{\tilde g}}\over{m_{\rm top}}}S_{\tilde g}\cr
A=&A_t+{{m_{\tilde g}}\over{m_{\rm top}}}A_{\tilde g}\cr
}$$
$$\eqalignno{
S_t=&K_{11}\lbrack c^2_\Theta C_1^{\tilde b_1\tilde t_1}
+s^2_\Theta C_1^{\tilde b_1\tilde t_2}\rbrack
+K_{21}\lbrack s_\Theta c_\Theta(C_1^{\tilde b_1\tilde t_1}
-C_1^{\tilde b_1\tilde t_2})\rbrack\cr
A_t=&S_t\cr
S_{\tilde g}=&K_{11}\lbrack c_\Theta s_\Theta
(C_0^{\tilde b_1\tilde t_2}-
C_0^{\tilde b_1\tilde t_1})\rbrack-K_{21}\lbrack c_\Theta^2
C_0^{\tilde b_1\tilde t_2}
+s^2_\Theta C_0^{\tilde b_1\tilde t_1}\rbrack\cr
A_{\tilde g}=&S_{\tilde g}\cr
K_{11}=&1-{{m^2_W}\over{m^2_{\rm top}}}\tan\beta\sin 2 \beta \cr
K_{21}=&{1\over{ m_{\rm top}}}(A_{\rm top}+\mu\tan\beta)\cr
C_0^{\tilde b_j\tilde t_i}=&-\int\limits_0^1d\alpha_1
\int\limits_0^{1-\alpha_1}d\alpha_2
{{m^2_{\rm top}}\over {f_{\tilde g}
^{\tilde b_j\tilde t_i}}}\cr
C_1^{\tilde b_j\tilde t_i}=&-\int\limits_0^1d\alpha_1
\int\limits_0^{1-\alpha_1}d\alpha_2
{{m^2_{\rm top}\alpha_1}\over{ f_{\tilde g}
^{\tilde b_j\tilde t_i}}}\cr
f_{\tilde g}^{\tilde b_j\tilde t_i}=&m^2_{\tilde g}-(m^2_{\tilde g}-
m_{\tilde t_i}^2)\alpha_1-(m^2_{\tilde g}-m^2_{\tilde b_j})\alpha_2-
m^2_{\rm top}\alpha_1(1-\alpha_1-\alpha_2)
-m^2_{H^+}\alpha_1\alpha_2\cr}
$$
where $c_\Theta=\cos\Theta$\ and $s_\Theta=\sin\Theta$.
S and A indicate that the contribution
comes from the scalar and axial scalar coupling
of the matrix element. For no mixing of the scalar
top quark masses the contribution of $S_{\tilde g}$\
and $A_{\tilde g}$\ are zero.
The Feynman integration can be done numerically.
\hfill\break\indent
\hfill\break\indent
To compare the standard QCD correction given in ref.6 with the
gluino and scalar quarks contribution
I present in Fig.1 and Fig.2
the results for different masses of the gluino and $\tan\beta$.
I take $\mu=500$ GeV and $A_{\rm top}=m_S$.
In the MSSM we have $m_{H^+}^2=m_W^2+m_{H_3^0}^2$\ where
$H^0_3$\ is the pseudo Higgs particle. That is the mass of the
charged Higgs particle has to be larger than the mass of the
W boson.
I put the charged Higgs mass to be equal $m_W$\ and the
top quark mass to be the recently released CDF value of
 174 GeV$^{13}$.
\hfill\break\indent
In Fig.1 and Fig.2 the solid straight line is the standard
contribution given in ref.6 and lies at $-9.5$\%.
I present the results for three different values
of the gluino mass that is 3 GeV (solid line), 100 GeV (dotted
line) and 500 GeV (dash-dotted line).
\hfill\break\indent
\vfill\break
\phantom{a)}
\vskip5.5cm
$$\vbox{\settabs2\columns\tenrm
\+ Fig.1: The ratio of $\Gamma^1/\Gamma^0$\ as a function
&\quad
Fig.2: The same as Fig.1 with \cr
\+ of the scalar mass $m_S$\ as explained in
&\quad $\tan\beta=10$\cr
\+ text.&\quad \cr}$$\indent
In Fig.1 I consider the case $v_1=v_2$. The lighter scalar
top quark mass is about 250 GeV for $m_S$\ smaller than 100 GeV,
decreases constantly to about 70 GeV for $m_S=350$\ GeV and
increases again to 260 GeV in the range considered here.
The heavier one varies from 358 GeV to 631 GeV
and $m_{\tilde b_{1,2}}=m_S$. Here
$\cos\Theta=1/\sqrt{2}$\ and the influence of the scalar
and axial scalar coupling of the gluino is larger than
the one of the top quark.
As a result we have in this case that the standard QCD corrections
are diminished for small gluino masses whereas we get an
enhancement up to $-18$\% for a gluino mass of 500 GeV.
Changing the $\mu$- parameter hardly affects the results.
If the Higgs mass
is enhanced all curves are pushed up closer to 0, but the
shape of the curves remains the same. For $m_{H^+}=120$\ GeV
the standard QCD correction is about $-8.1$\%.
\hfill\break\indent
In Fig.2 I consider the same case, but with $\tan\beta=10$.
Here the contribution of the gluino are much larger
than the top quark contribution due to a $\mu\tan\beta$\
dependence in the couplings.
The gluino mass also
becomes more important, wheras for very large gluino masses
($m_{\tilde g}
\gg 100$\ GeV) the 1 loop contribution $\Gamma^1(t\rightarrow
H^+b)$\ is decreasing again.
Here $\cos\Theta\approx 1/\sqrt{2}$,
the lighter scalar top quark mass is about 115 to 110 GeV for
$m_S$\ smaller than 100 GeV and increases constantly to 379 GeV
for $m_S=450$\ GeV.
The heavier one varies from 219 GeV to 564 GeV.
The heavy scalar bottom quark mass varies from 78 GeV
to 454 GeV and the lighter one from 56 GeV to 451 GeV.
As a result we see that the gluino mass contribution
enhances the standard QCD correction drastically.
This decay
mode therefore can be used to put constraints on the ratio
of the vevs $v_1$\ and $v_2$.
Smaller values for $\mu$\ diminishes the results whereas
higher values for $\mu$\ enlarges them. For $m_S\leq 50$\
GeV the ratio $\Gamma^1/\Gamma^0$\ is decreasing again.
\vglue 0.2cm
{\elevenbf III. FINAL REMARKS AND CONCLUSIONS}
\vglue 0.2cm
$v_1\ll v_2$\ has to be taken with care, because we neglected
the mass of the bottom quark. In a full analysis I also
have to include the scalar partner of the right handed
bottom quark, whose coupling are proportional to the bottom quark
mass. Although if I take $\tan\beta=2$, that is
$m_b\tan\beta\ll m_{\rm top}\cot\beta$, the shape of the
curves remains the same as in Fig.2, but with values closer to
the standard model.\hfill\break\indent
Furthermore the most competitive
decay mode to $t\rightarrow H^+b$\ is the equivalent decay mode
of $t\rightarrow W^+b$, which was considered in ref.14--16.
There it was shown that the corrections are in the same order
as for the $t\rightarrow H^+b$\ decay, that is $-10$\% according
to the equivalence theorem. In ref.6 it was also shown
that the $t\rightarrow W^+b$ decay is becoming more
important than the $t\rightarrow H^+b$\ for increasing
$\tan\beta$.\hfill\break\indent
In ref.2 and ref.3 it was shown that the electroweak corrections
to $t\rightarrow H^+b$\ are in the order of $-5$\% for a heavy
top quark and decreases rapidly for higher values of
$\tan\beta$.\hfill\break\indent
Last but not least
 within the MSSM the electroweak corrections
to the $t\rightarrow W^+b$\ decay mode was recently
 considered in ref.17. They considered the contribution
of neutralinos and charginos to this decay mode. They
showed that the contributions are of the order of
$-5$\% to $-10$\% and increasing for higher values of
$\tan\beta$.\hfill\break\indent
In this talk I have shown that in a full analysis of the
$t\rightarrow H^+b$\ decay one cannot neglect the gluino
and scalar quark masses. As a result, because of the
equivalence theorem,
the same must be true
for the $t\rightarrow W^+b$ decay.
A full analysis
of the gluino and scalar quarks
contribution to this decay rate has not been done
yet and will be presented elsewhere$^{18}$.
\vglue 0.52cm
{\elevenbf IV. ACKNOWLEDGMENTS}
\vglue 0.52cm
I would like to thank the physics department
of Carleton university for the use of their computer
facilities. The figures were done with the very user
--friendly program PLOTDATA from TRIUMF.
\hfill\break\indent
This work was partially funded by funds from the N.S.E.R.C. of
Canada and les Fonds F.C.A.R. du Qu\'ebec.
\vglue 0.52cm
{\elevenbf REFERENCES}
\vglue 0.52cm
\item{[\ 1]}C.S. Li, B.Q. Hu and J.M. Yang, {\elevenit Phys.Rev.}
{\elevenbf D47}(1993)2865.
\item{[\ 2]}A. Czarnecki,
 {\elevenit Phys. Rev.}{\elevenbf D48}(1993)5250
\item{[\ 3]}C.S. Li and T.C. Yuan,
{\elevenit Phys.Rev.}{\elevenbf D42}(1990)3088,
erratum-ibid{\elevenbf D47}(1993)2556.
\item{[\ 4]}C.S. Li, Y.S Wei and J.M. Yang,
 {\elevenit Phys. Lett.}{\elevenbf B285}(1992)137.
\item{[\ 5]}J. Liu and Y.P. Yao, {\elevenit Phys.Rev.}
{\elevenbf D46}(1992)5196.
\item{[\ 6]}A. Czarnecki and S. Davidson,
 {\elevenit Phys.Rev.}{\elevenbf D48}
(1993)4183, {\elevenit Phys.Rev.}{\elevenbf D47}(1993)\
3063.
\item{[\ 7]}J.F. Gunion et al.,
"{\elevenit The Higgs Hunter's Guide}"
(Addison-Wesley, Redwood City, CA, 1990).
\item{[\ 8]}H.P. Nilles, {\elevenit Phys.Rep.}{\elevenbf 110}(1984)1.
\item{[\ 9]}H.E. Haber and G.L. Kane,
{\elevenit Phys.Rep.}{\elevenbf 117}(1985)75.
\item{[10]}H. K\"onig, "{\elevenit QCD corrections to the $t\rightarrow
H^+b$\ decay within the minimal supersymmetric standard
model}", UQAM-PHE-94/01, hep-ph/9403297.
\item{[11]} see e.g. HELIOS collaboration, T. Akesson et al,
{\elevenit Z.Phys.}{\elevenbf C52}(1991)219 and references therein.
\item{[12]}J. Ellis, D.V. Nanopoulos and D.A. Ross,
{\elevenit Phys.Lett.}
{\elevenbf B305}(1993)375.
\item{[13]}CDF Collaboration, Fermilab preprint, April 1994.
\item{[14]}M.Jezabik and J.H. K\"uhn,
{\elevenit Nucl.Phys.}{\elevenbf B314}(1989)1,
ibid {\elevenbf B320}(1989)20.
\item{[15]}J.Liu and Y.P.Yao, UM-Tth-90-09 (1989), {\elevenit Phys.Rev.}
{\elevenbf D46}(1992)467.
\item{[16]} A. Czarnecki, {\elevenit Phys.Lett.}{\elevenbf B252}(1990)467.
\item{[17]}D. Garcia, R.A. Jim\'enez, J. Sol\`a and W. Hollik,
"{\elevenit Electroweak supersymmetric Quantum corrections to the top
quark decay}", UAB-FT-323, hep-ph/9402341.
\item{[18]}H. K\"onig, in preparation.
\hfill\break\vskip.12cm\noindent
\vfill\break
\eject\bye